\documentclass[aps,pre,reprint,twoside,showkeys,longbibliography,superscriptaddress]{revtex4-2}

\usepackage{amsmath}
\usepackage{amsfonts}
\usepackage{amssymb}
\usepackage{graphicx}
\usepackage{dcolumn}
\usepackage{bm}
\usepackage{xcolor}
\usepackage{url}
\usepackage{amssymb}
\usepackage{amsmath}
\usepackage[hidelinks]{hyperref}
\usepackage[caption=false]{subfig}
\usepackage{booktabs}
\usepackage{placeins}
\usepackage{todonotes}

\newcommand{\Penalty}{\Theta}
\newcommand{\Penaltycomm}{\theta}
\newcommand{\Admissible}{\mathcal{A}}
\newcommand{\nmin}{n_\text{min}}
\newcommand{\nmax}{n_\text{max}}
\newcommand{\commconstraintweight}{\phi}

\DeclareMathOperator*{\argmaxB}{\arg\max} 


\begin{document}


\title{Heuristic and exact modularity optimization with size-constrained communities}

\author{Filipi N. Silva}
\affiliation{Kellogg School of Management, Northwestern University, Evanston, IL, USA}

\author{Samin Aref}
\affiliation{Department of Mechanical and Industrial Engineering, University of Toronto, Toronto, ON, Canada}

\author{Vincent Traag}
\affiliation{Centre for Science and Technology Studies (CWTS), Leiden University, the Netherlands}

\author{Santo Fortunato}
\affiliation{Center for Complex Networks and Systems Research, Luddy School of Informatics, Computing, and Engineering, Indiana University Bloomington, USA}

\date{\today}

\begin{abstract}

When searching for communities in networks, domain experts may have some prior expectations about the size of communities. Yet, community detection methods normally do not optimize communities under cluster size constraints. Multi-resolution techniques allow users to indirectly control the average community size through changing a resolution parameter, but this practice does not control the size of individual communities. We here study the problem of size-constrained community detection, where the size of all communities is limited to a user-specified range of values, in the context of modularity optimization.
We propose a heuristic for modularity optimization under community size constraints. To demonstrate the reliability of our proposed heuristic, we also formulate an exact integer optimization model and use its results as a baseline. Our analysis based on synthetic benchmarks and real networks demonstrate the issues with the currently common practice of changing resolution parameters and reveal the advantages of the proposed methods as a principled way of obtaining size-constrained communities. The proposed method is publicly available in the Python Leiden algorithm package.
\end{abstract}

\maketitle


\section{Introduction}
\label{sec:introduction}

Understanding the underlying organizational principles of networks is a fundamental endeavor, and community detection has emerged as a pivotal concept in this field. Communities within networks represent subsets of nodes that are relatively well-connected internally~\cite{fortunato10}.

Community Detection is a fundamental network problem which, in its simplest definition, involves partitioning nodes of an input network into communities according to the network structure. The identification of these communities is a crucial stepping stone toward comprehending the intricate structure, information flow, and emergent behavior within complex systems. 

Community detection is applied in various settings, spanning disciplines such as sociology, biology, and computer science. Practitioners in these fields apply community detection algorithms to analyze network models which often represent field-specific interconnected systems as the subject of a study. Despite the variety of field-specific contexts and applications, researchers in different fields may have a common computational challenge that is yet to be addressed: The subject matter expert may have prior information or requirements on community sizes. However, most existing community detection algorithms do not offer the user the flexibility to specify and directly control minimum and maximum community sizes.

One example that shows the relevance of community size constraints appears in a common business application of community detection: market segmentation. This application involves clustering nodes that represent customers into communities using their networked data or other data purchasing history \cite{korczak2019approach,zhang2021measuring}. 
A common issue with this usage is that communities will rarely become practical market segments because existing unconstrained community detection methods often lead to one extremely large community of many customers and numerous small communities of very few customers \cite{korczak2019approach,zhang2021measuring}. Most small communities are often discarded without further analysis \cite{zhang2021measuring} because they contain too few customers to represent meaningful segment needs or be worthwhile for targeted offerings \cite{korczak2019approach}. However, the range of acceptable sizes for market segments (community sizes) may be available based on marketing subject matter expertise. Therefore, a community detection algorithm that incorporates such community size range will be particularly useful in this context.

The same challenge appears in many contexts across different fields including neuroscience applications where community detection algorithms are applied on brain connectome networks \cite{Budapest2015}. The spatial and wiring patterns of neural links in brain connectome networks lead to existing algorithms typically producing either two or four trivial communities corresponding to two brain hemispheres or the front and back split of the two hemispheres, which are not particularly informative from a neuroscience perspective.


Yet another example showing the same common challenge is chromosomal interaction networks. Unconstrained modularity-based community detection for the chromosomal interaction network in \cite{mokhtaridoost2024} leads to a 3-community partition, whose communities are so large that they contradict exogenous information from biology including \textit{chromosomal territories} \cite{chromosomal} and \textit{topologically associating domain} boundaries \cite{topologically}. Therefore, size-constrained community detection will be particularly useful in genome topology research, where field-specific knowledge of reasonable ranges of community sizes is often available \cite{mokhtaridoost2024,chromosomal,topologically}.

We here study this problem further, focusing in particular on modularity~\cite{newman_finding_2004} as a commonly used and well studied method for community detection. Multi-resolution modularity-based algorithms allow users to change the modularity resolution parameter in an ad-hoc way to obtain communities matching their community size requirements using trial and error. Our results show that this common approach has major disadvantages, while its common usage justifies the need for a principled method for size-constrained community detection. We propose a modularity-based heuristic that incorporates community size constraints. We also formulate an exact method and use its results as baseline to demonstrate the reliability of our proposed optimization despite its heuristic nature \cite{aref2024analyzing}.

\subsection{Preliminaries and Notation}

We represent the undirected and unweighted graph $G$ as the pair of sets $V$ and $E$, $V$ being the set of nodes ($|V|=n$), and $E$ the set of edges ($|E|=m$). The symmetric adjacency matrix of graph $G$ is represented by $\textbf{A}=[a_{ij}]$. The degree of node $i$ is represented by $d_i=\sum_j a_{ij}$. 

Given a partition $P=\{V_1,V_2, \dots, V_q \}$ of the node set $V$ into $q$ non-overlapping communities, the modularity function $Q(G,P)$ is computed as~\cite{newman_modularity_2006,fortunato2016} 

\begin{equation}
\label{eq0}
 Q(G,P)= \frac{1}{2m} \sum \limits_{(i,j) \in V^2} \left( a_{ij} - \gamma\frac{d_id_j}{2m}\right) \delta(i,j)\,.
\end{equation}

In Eq.\ \eqref{eq0}, $\gamma$ is the resolution parameter. The Kronecker delta, $\delta(i,j)$, equals 1 if nodes $i$ and $j$ are in the same community, otherwise it equals 0.

The symmetric modularity matrix of graph $G$ for a given resolution value $\gamma$, is represented by $\textbf{B}=[b_{ij}]$, whose entries are $b_{ij} = a_{ij} - \gamma d_{i}d_{j}/{2m}$.
Some ordered pairs of nodes $(i,j)$ correspond to non-negative modularity entries $b_{ij} \geq 0$; the set $B^+=\{(i,j) \in V^2 , i < j \mid b_{ij} \geq 0\}$ represents all such ordered pairs. Conversely, the set $B^- = \{(i,j) \in V^2 , i < j \mid b_{ij} < 0\}$ contains all the ordered pairs of nodes that correspond to negative modularity entries $b_{ij} < 0$.

\section{Problem statements}

In this section, we provide four problem statements to better define the optimization tasks involved in this study.

\begin{itemize}
    
\item The unconstrained maximum modularity partition (MM) problem for the input graph $G=(V,E)$ involves finding a partition $P^*(G)$ whose modularity is maximal over all possible partitions: $P^*(G)=\argmaxB_P Q(G,P)$.

\item The size-constrained maximum modularity partition (SMM) is the same as the MM problem, but it is subject to the community size constraints: $\nmin \leq n_c \leq \nmax$, $\forall c$, where $n_c$ is the size of community $c$. The set of partitions $P$ that satisfy the above constraints is denoted by $\Admissible$.

\item The maximum modularity $k$-partition (MMk) problem for the input graph $G=(V,E)$ involves optimizing the modularity function $Q(G,P)$ over all possible partitions of nodes to $k$ communities at most. 

\item The size-constrained maximum modularity $k$-partition (SMMk) problem is the same as the MMk problem, but it is subject to the community size constraints: $\nmin \leq n_c \leq \nmax$, $\forall c$.

\end{itemize}

\section{Exact optimization baseline}
\label{s:exact}

The MM problem is well studied and several integer programming (IP) formulations exist for it \cite{brandes2007modularity,agarwal_modularity-maximizing_2008,dinh_toward_2015,aref2023suboptimality,aref2022bayan} but community size constraints cannot be directly added to them.  
Therefore, we formulate the MMk problem as an integer program and then further extend it to represent the SMMk problem. We then use the exact optimization results from the SMMk as baseline for assessing the reliability of the optimization process in our proposed heuristic method.

\subsection{Formulating the MMk problem}

First, we formulate the MMk problem as an Integer Program (IP) in Eqs.~\eqref{eq:generic_ip_obj}--\eqref{eq:generic_ip_constraints} where objective function and constraints are all linear. We formulate the objective function of the MMk problem in Eq.\ \eqref{eq:generic_ip_obj}. The binary decision variable $f_{ij}$ takes the value 1 if and only if nodes $i$ and $j$ belong to the same community in the corresponding solution. The objective function is the modularity of the partition corresponding to the IP solution. Therefore, the optimal value of the objective function equals the maximum modularity of the input graph $G$ over partitions into at most $k$ communities ($k$-partitions), with $k \leq n$ chosen by the user. 

\begin{equation}  
\label{eq:generic_ip_obj}
 \begin{aligned} 
 \max_{x_{ij},f_{ij}} Q &= \frac{1}{2m} \left( \sum\limits_{(i,j) \in V^2 , i< j} 2b_{ij}f_{ij} + \sum\limits_{(i,i) \in V^2} b_{ii} \right) \\ 
 \end{aligned} 
\end{equation}

The constraints are formulated in Eq.\ \eqref{eq:generic_ip_constraints} such that the formulation becomes a linear model (because integer \textit{linear} programs can be solved efficiently using mathematical solvers). The binary decision variable $x_{ic}$ takes the value 1 if and only if node $i$ belongs to community $c$. In Eq.\ \eqref{eq:generic_ip_constraints}, the set $C=\{0,1,2,\dots,k-1\}$ contains the $k$ potential community indices. Some of the $k$ potential communities may be empty in the partition obtained from the solution of the IP problem.  
The first constraint in Eq.\ \eqref{eq:generic_ip_constraints} ensures that node $i$ belongs to precisely one community in every feasible solution. The second and third constraints ensure that $f_{ij}$ takes the correct value for the pair of nodes $(i,j)$ that belongs to $B^+$. These two constraints are redundant when nodes $i,j$ are assigned to the same community. However, when $i,j$ are assigned to different communities, one of the two constraints activates and forces $f_{ij}$ to be 0 (against the pressure of the objective function). The fourth constraint ensures that $f_{ij}$ takes the correct value for the pair of nodes $(i,j)$ that belong to $B^-$. This constraint is redundant when nodes $i,j$ are assigned to different communities. However, it activates when nodes $i,j$ are assigned to the same community and forces $f_{ij}$ to be 1 (against the pressure of the objective function). 

\begin{equation}  
\label{eq:generic_ip_constraints}
 \begin{aligned} 
 \quad \sum _{c \in C} x_{ic}&= 1; \quad \forall i \in V \\
 f_{ij}&\le 1 - (x_{ic} - x_{jc}); \quad \forall (i,j) \in B^+, ~\forall c \in C \\
  f_{ij}&\le 1 - (x_{jc} - x_{ic}); \quad \forall (i,j) \in B^+, ~\forall c \in C \\
 f_{ij}&\ge x_{ic} + x_{jc} -1; \quad \forall (i,j) \in B^-,
 ~\forall c \in C \\ x_{ic}&\in \{0,1\}; \quad \forall i \in V, ~\forall c \in C \\ 
 f_{ij}&\in \{0,1\}; \quad \forall (i,j) \in V^2,\, i< j \,.
 \end{aligned} 
\end{equation}

To obtain exact optimization results on the MMk problem, the model in Eqs.~\eqref{eq:generic_ip_obj}--\eqref{eq:generic_ip_constraints} can be solved to global optimality for small input graphs using an IP solver like Gurobi \cite{gurobi}. Alternatively, the model in Eqs.~\eqref{eq:generic_ip_obj}--\eqref{eq:generic_ip_constraints} can be solved with the positive \textit{mixed integer programming gap} percentage of $0<\varepsilon<1$ using Gurobi to obtain a partition within $\varepsilon$ percent of global optimality. An optimal solution of the model in Eqs.~\eqref{eq:generic_ip_obj}--\eqref{eq:generic_ip_constraints} represents a maximum-modularity $k$-partition for the input graph $G$. For sufficiently large values of $k$ (e.g., $k=n$), an optimal solution of the model in Eqs.~\eqref{eq:generic_ip_obj}--\eqref{eq:generic_ip_constraints} represents a maximum-modularity partition for the input graph $P^*(G)=\argmaxB_P Q(G,P)$.

\subsection{Formulating the SMMk problem}
The community size constraints require the size (number of nodes) of all communities to be within a predefined range: $\nmin \leq n_c \leq \nmax$, $\forall c$. The IP formulation for the SMMk problem includes Eqs.~\eqref{eq:generic_ip_obj}--\eqref{eq:generic_ip_constraints} (has the same decision variables and objective function), but it also has additional constraints and therefore a more restricted feasible space.

Given that the IP in Eqs.~\eqref{eq:generic_ip_obj}--\eqref{eq:generic_ip_constraints} has cluster indices based on the potential (and not the actual) number of communities, enforcing the $\nmin$ constraint requires several constraints. We include $kn$ additional linear constraints as defined in Eq.\ \eqref{eq:nmin} to ensure that all non-empty community sizes are at least $\nmin$.

\begin{equation}
\label{eq:nmin}
    \sum _{i \in V} x_{ic} \geq \nmin x_{ic}; \quad \forall i \in V, \quad \forall c \in C\,.
\end{equation}

Enforcing the $\nmax$ constraint is easier because we do not need to separate the empty from non-empty communities among the $k$ potential communities. We include $k$ additional constraints as defined in Eq.\ \eqref{eq:nmax} to ensure that all community sizes are at most $\nmax$.

\begin{equation}
\label{eq:nmax}
    \sum _{i \in V} x_{ic} \leq \nmax;  \quad \forall c \in C\,.
\end{equation}

The IP formulation for the SMMk problem includes Eqs.~\eqref{eq:generic_ip_obj},\eqref{eq:generic_ip_constraints},\eqref{eq:nmin}, and \eqref{eq:nmax}. This IP formulation can be solved exactly (or approximately) by Gurobi to obtain (the approximation of) a globally optimal $k$-partition for the input graph under the community size range constraint of $\nmin \leq n_c \leq \nmax$.

\section{Heuristic optimization method}
\label{s:heuristic}

Directly enforcing the community size constraints in modularity-based heuristic algorithms is not straightforward.
For example, if $\nmin > 1$, starting from a singleton partition is not an admissible solution for heuristics that rely on it.
Hence, if we want to enforce these constraints directly, we should start with finding an alternative initial solution that remains admissible when the community size range is enforced.
Even if we would be able to come up with a trivial initial admissible solution, it might not be straightforward to change this initial solution while enforcing the constraints.
Consider the example of a 6-node line graph (6 nodes connected in a line of 5 edges) that is $a\text{--}b\text{--}c\text{--}d\text{--}e\text{--}f$.
Suppose we have community size constraints of $\nmin = 2$ and $\nmax = 4$, and we have an initial partition that consists of two communities: $V_1 = \{a, b, c\}$ and $V_2 = \{d, e, f\}$.
Perhaps, given the particular quality function, it would favor the partition of three communities of $V'_1 = \{a, b\}$, $V'_2 = \{c, d\}$ and $V'_3 = \{e, f\}$.
Starting from the initial partition $V_1, V_2$, it would be admissible to move node $c$ to community $V_2$. However, it would not be admissible to move node $c$ to a new community $V'$, to which then also node $d$ could be moved to, since community $V'$ would be initially of size $|V'| = 1 < \nmin$.
Hence, enforcing constraints directly does not seem a viable route given the local moves of nodes that are common in modularity-based heuristics.

We instead propose to transform the constrained problem into an unconstrained optimization problem.
There are two broad approaches for this conversion: a barrier approach or a penalty approach.
The barrier approach ensures that the optimization always stays within the boundaries of the constraints. For this reason, the barrier approach is also known as the interior-point method.
Although this approach works well for cases such as linear programming \cite{karmarkar1984new}, it does not work well in our case.
This is because heuristic algorithms may benefit from temporarily moving outside the constraints' boundaries to explore potentially better partitions.

We focus on using the penalty method for converting the constrained problem into an unconstrained problem.
Instead of optimizing $Q(G,P)$, we will optimize $Q(G,P) - \commconstraintweight\Penalty(P)$, where $\commconstraintweight$ is a weight for how strongly the constraints $\Penalty(P)$ should be penalized. 
To set $\commconstraintweight$, we use a progressive schedule.
We begin with the unconstrained problem ($\commconstraintweight=0$).
Whenever the resulting partition violates the community size bounds, we increase $\commconstraintweight$, starting from $10^{-4}$ and doubling it at each step.
Each optimization is initialized from the partition found at the previous value of $\commconstraintweight$.
Hence, the final solution corresponds to the first penalty level that yields a feasible partition.

The $\Penalty(P)$ should always be finite, for any partition $P$, otherwise we end up using a barrier approach.
Within the space of admissible solutions, the penalty should play no role, and hence $\Penalty(P) = 0$ for $P \in \Admissible$, where $\Admissible$ is the space of all admissible partitions, while $\Penalty(P) > 0$ for $P \not\in \Admissible$.
We formulate $\Penalty$ to be a sum over the different communities, that is
$$\Penalty(P) = \sum_c \Penaltycomm(c)$$
where $\Penaltycomm$ is a penalty function for a particular community.

In addition, the penalty $\Penaltycomm$ should be sub-additive.
We argue why this is necessary using a counterexample 
Assume $\Penaltycomm$ is super-additive, so that $\Penaltycomm(a + b) \geq \Penaltycomm(a) + \Penaltycomm(b)$.
As a simple counterexample, consider an empty graph of size $n$ with a minimum community size constraint of $n$, that is we want to find only a single community.
To further simplify the counterexample, let us assume we are optimizing modularity with a resolution parameter $\gamma = 0$.
From the perspective of modularity $Q(G,P)$, then $Q(G,P) = 0$ for any partition $P$ since the community sizes play no role and there are no edges to consider.
However, $\Penalty(P)$ will play a role.
Consider a partition $P$ of two communities of size $\frac{n}{2}$ each. 
Clearly, this does not satisfy the community size constraint of $n_c \geq n$.
To move towards a partition that consists of a single community, we may try to move nodes from one cluster to another.
We thus arrive at a partition $P'$ with community sizes $n_1 = \frac{n}{2} - 1$ and $n_2 = \frac{n}{2} + 1$.
Assuming that $\Penaltycomm$ is super-additive, we arrive at
\begin{equation}
\begin{split}
\Penaltycomm\left(\frac{n}{2} + 1 \right)
+ \Penaltycomm\left(\frac{n}{2} - 1 \right)
\geq \Penaltycomm\left(\frac{n}{2}\right) \\
\geq 2\Penaltycomm\left(\frac{n}{2}\right)
+ \Penaltycomm(1) + \Penaltycomm(-1) \\
\geq 2\Penaltycomm\left(\frac{n}{2}\right)
\end{split}
\end{equation}
Hence $Q(G,P') - \Penalty(P') \leq Q(G,P) - \Penalty(P)$ so that partition $P$ is preferred over $P'$.
In a heuristic algorithm moving individuals nodes, we would hence tend to stay near an equisized partition $P$ instead of moving towards the partition with a single community.
One choice that satisfies these constraints is
\begin{equation}
    \Penaltycomm(n_c) = 
        \begin{cases}
            \sqrt{\nmin - n_c}, & \text{if~} n_c < \nmin\,; \\
            \sqrt{n_c - \nmax}, & \text{if~} n_c > \nmax\,. \\
        \end{cases}
\end{equation}

\section{Results}

In this section, we provide comparative experiments using three methods: (1) Unconstrained Leiden - UL (to represent an unconstrained modularity heuristic), (2) Constrained Leiden - CL (implementation of the penalty approach described in Section \ref{s:heuristic}), and (3) Constrained Integer Programming - CIP baseline (implementation of the exact and approximation approach described in Section \ref{s:exact}). Among the performance measures reported for each method, we use the Adjusted Mutual Information (AMI). AMI takes a partition obtained by an algorithm and a reference partition (e.g., the planted partition of a synthetic benchmark network) and returns a value from the unit interval indicating the similarity between the two partitions \cite{vinh_AMI}. Compared to the normalized mutual information, AMI is a more reliable measure of partition similarity \cite{jerdee2025normalized}.

\subsection{Detailed analysis of an illustrative example}
\label{ss:toy}

\begin{figure}
    \centering
    \includegraphics[width=0.98\linewidth]{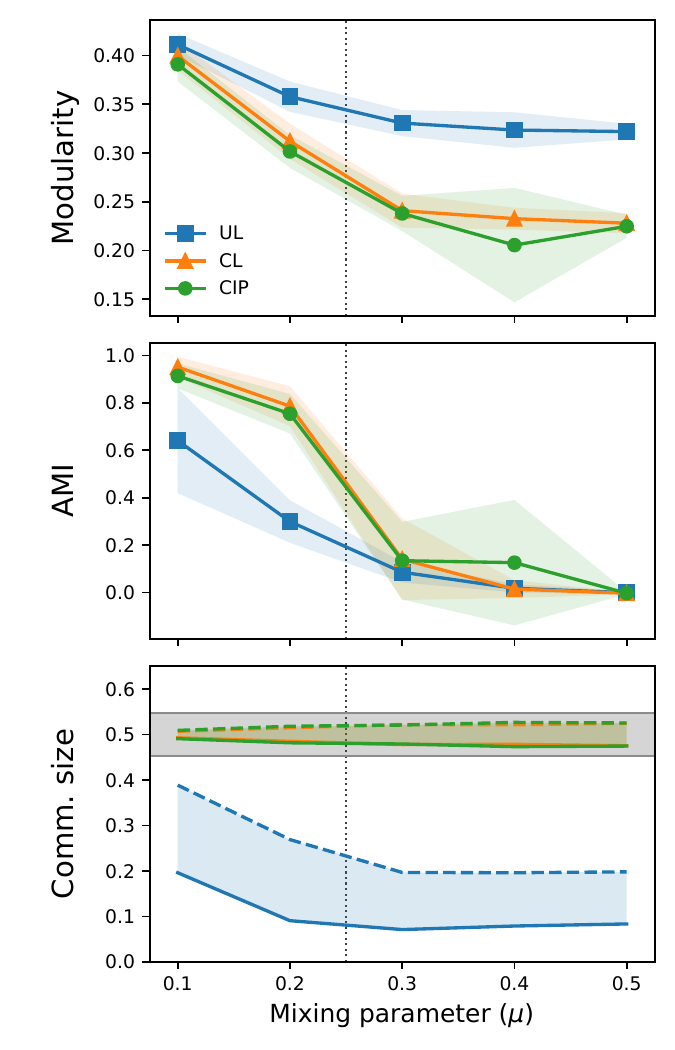}
  \caption{Tests of UL, CL, and CIP on the planted partition model. Panel (a) shows the standard modularity values of the partitions produced by the three methods; UL often attains higher standard modularity because it is not subject to community-size constraints. Panel (b) shows that the planted partitions are retrieved at considerably higher rates by CL and CIP compared to UL for small and medium values of $\mu$. Panel (c) shows that UL's comparative lower retrieval rate is due to community sizes which are not consistent with the range of community sizes in the planted partition model. CL and CIP consistently produce communities that are within the expected range of sizes. The vertical dotted line marks the detectability limit for this configuration.}
    
    \label{fig:toy_example}
\end{figure}

As an illustrative example, we use a series of planted partition graphs~\cite{condon01} with $n=128$, average degree $k=8$, two communities, and varying mixing parameter $\mu$. In this model, nodes are assigned to predefined communities and edges are generated  with different probabilities for intra- and inter-community connections. The mixing parameter controls the fraction of inter-community edges, and larger values of $\mu$ reduce the contrast between groups, making the community detection task harder. The community size range considered is $[n/2 - n/2\times 10\%, n/2 + n/2\times 10\%]$.

Figure \ref{fig:toy_example} illustrates the differences between the three approaches UL, CIP, and CL. The partitions found by UL were observed to consistently violate the community size range while CIP and CL both satisfy the community size constraint across all values of the mixing parameter $\mu$ as shown in Figure \ref{fig:toy_example}(c). CIP produces a wider range of community sizes compared to CL. CIP's wider range of community sizes is expected because the community size IP constraints are satisfied marginally to squeeze all the potential gains for the objective function. 

Figure \ref{fig:toy_example}(b) shows that for UL, the violation of the community size constraints corresponds to substantially lower retrieval rates (AMI) compared to the two other methods for small and medium values of $\mu$ that are below the \textit{detectability limit} \cite{ghasemian2016detectability}. The distinct behavior from these three methods cannot be observed by the modularity of their partitions alone. Figure \ref{fig:toy_example}(a) shows that UL often attains higher standard modularity than the two constrained methods, illustrating the trade-off between unconstrained modularity optimization and satisfying the desired community-size range. However, Figure \ref{fig:toy_example}(b) shows that CL, despite its heuristic nature, returns partitions with AMIs that are close or even equal to the AMIs of CIP.

Figure \ref{fig:toy_example} shows that, unlike UL, CL satisfies the imposed community size range, which in turn leads to a better retrieval of the planted partition. Also,  while the heuristic nature of CL may lead to globally suboptimal modularity values, it does not prevent CL from retrieving the planted partition as well as CIP. Taken together, we conclude that CL, despite its heuristic nature, is a reliable method for obtaining partitions within a desirable community size range.

\subsection{Planted partition benchmark networks}
\label{sec:modnet}

In this section, we provide results on the differences between the three methods, UL, CL, and CIP in retrieving planted partitions across a wider range of parameters for synthetic planted partition networks. We investigate 12 experimental settings that cover three values for the average degree $k$ and four values for the network size $n$. For each experimental setting, we create planted partition graphs using mixing parameters $\mu$ ranging from 0.1 to 1. In the figures, we show the range $\mu \leq 0.5$ because retrieval rates are near zero for larger values of $\mu$.

\begin{figure*}
    \centering
    \includegraphics[width=0.98\linewidth]{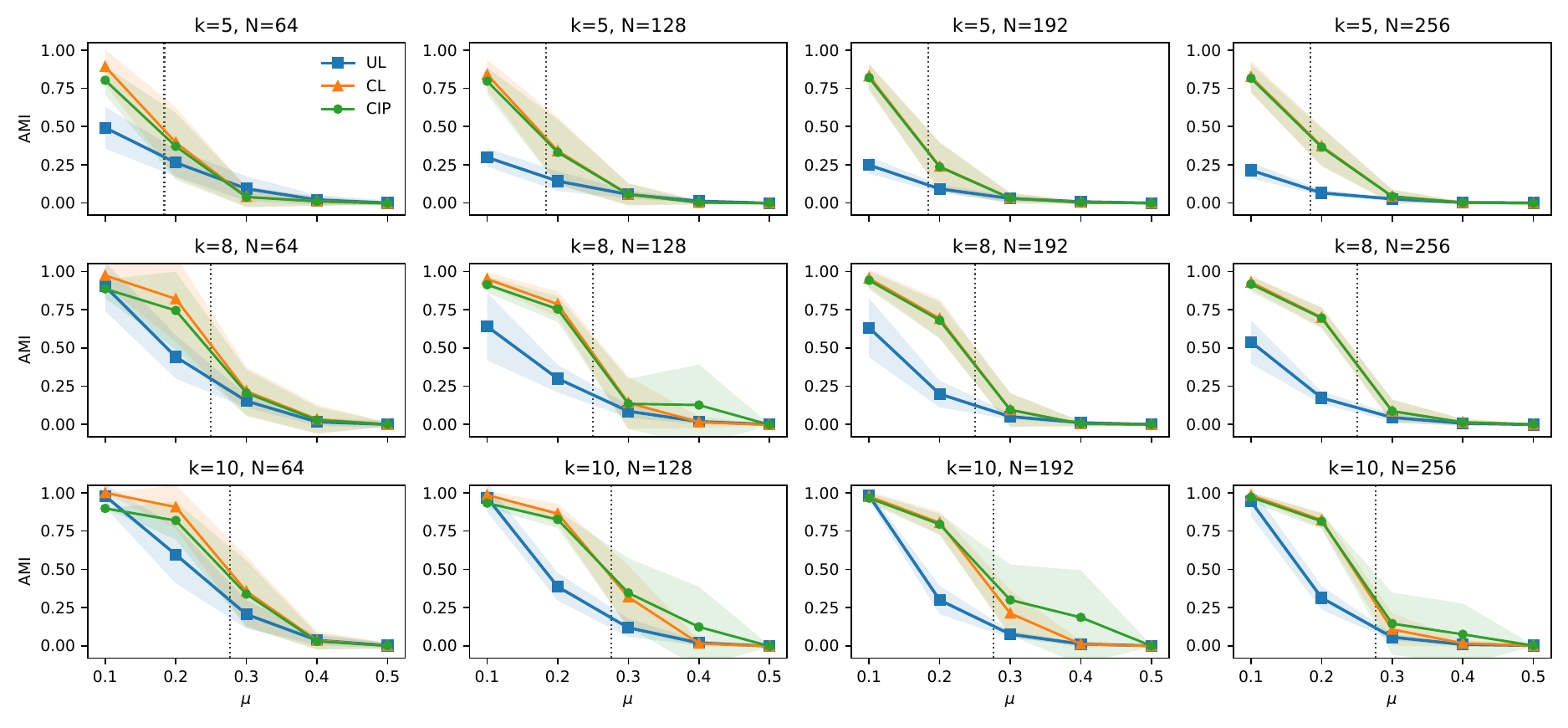}
    \caption{Retrieval rates (AMI) for the three methods UL, CL, and CIP on 12 experiment settings (each panel) of planted partition networks. Each row shows networks with the same degree, increasing from top to bottom from $k = 5$ to $k = 10$. Each column shows networks with the same number of nodes, increasing from $n = 64$ to $n = 256$. The vertical dotted line in each panel marks the detectability limit for the corresponding configuration.}
    \label{fig:ami_ppm}
\end{figure*}

Figure \ref{fig:ami_ppm} shows each experimental setting in a separate panel. All the three methods have very small AMI retrieval rates close to 0 for all networks with $\mu>0.4$ due to the detectability limit of the planted partition model \cite{ghasemian2016detectability}. Therefore, in our explanation of the results we focus on the results for $\mu \leq 0.4$ in Figure \ref{fig:ami_ppm}. We observe that the gap between UL and the two constrained methods in retrieval rate for networks with $\mu \leq 0.4$ slightly widens as the network order $n$ increases while $k$ remains constant. As the average degree $k$ increases for a fixed $n$, the retrieval rate gap between UL and the two other method shrinks. For networks with $k=5$ and $\mu \leq 0.4$, the retrieval rates of UL and that of the constrained methods have the largest difference.

Another noteworthy observation in Figure \ref{fig:ami_ppm} is the closeness of the retrieval rates for CL and CIP across all values of $n, k$, and $\mu$. This reaffirms the reliability of the optimization in CL despite its heuristic nature consistent with the illustrative example in Figure \ref{fig:toy_example}. CL retrieves planted partitions at rates equal or even higher than the IP-based constrained baseline in these experiments.

We observed the same general pattern of UL consistently producing communities outside of the desired range of communities sizes when we change the number of communities, number of nodes, and average degrees. Additional results on the standard modularity values of the partitions obtained by the three methods for the same planted partition networks are provided in Appendix A (Figure \ref{fig:modularity_constrained}). Additional results on the community sizes of the partitions found by the three methods for the same planted partition networks are provided in Appendix A (Figure \ref{fig:sizes_constrained}).

\subsection{Ring of cliques}

\label{sec:ringcliques}
\begin{figure}
    \centering
    \includegraphics[width=0.98\linewidth]{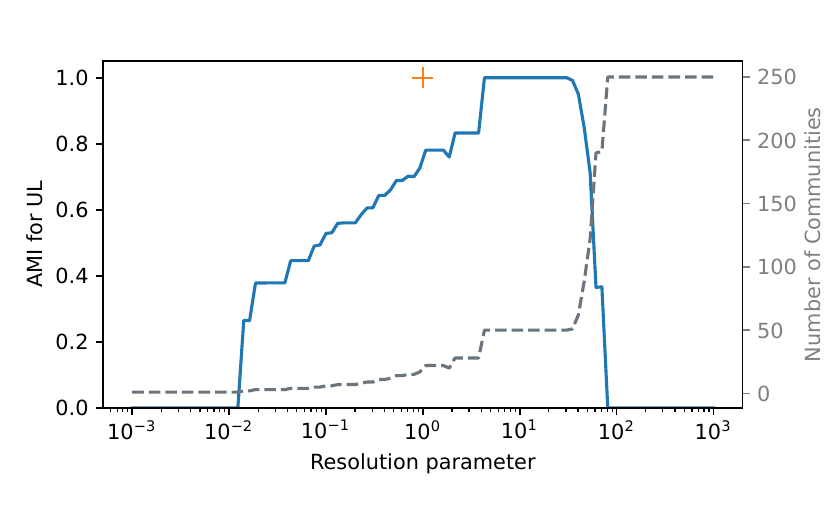}
    \caption{Correspondence (via AMI) between cliques and detected communities in a ring of cliques with $50$ cliques of size $5$ with pairs of consecutive cliques connected by a single edge. The cross marker indicates the value obtained by CL (using the known clique sizes) at resolution 1.0. The number of communities detected at each resolution parameter is shown as a dashed line.}
\label{fig:ring_results}
\end{figure}

We use the synthetic graph known as \textit{ring of cliques} to investigate the applicability of the CL method compared to the conventional practice of tuning the resolution parameter. The graph is made out of 50 cliques of 5 nodes, with consecutive cliques  connected to each other by a single edge. We consider the cliques as the planted communities. Figure \ref{fig:ring_results} shows the AMI for the single output partition of CL as well as the AMI for each output partition of UL corresponding to specific resolution parameter values. UL does not retrieve the planted partition unless the resolution parameter is within the range of $[5,40]$. In contrast, the orange cross marker in Figure \ref{fig:ring_results} shows that the CL method accurately retrieves the planted partition at the default resolution of 1 when run with community-size range of $[2, 8]$ (which includes the planted clique size of 5). Figure \ref{fig:ring_results} shows the number of communities on the right vertical axis indicating that resolution values below 5 lead to UL returning far fewer communities than 50 (where each community is comprised of multiple adjacent cliques).

\subsection{Brain connectome network}
\label{sec:brain} 

\begin{figure}[htbp!]
    \centering
    \includegraphics[width=0.8\linewidth]{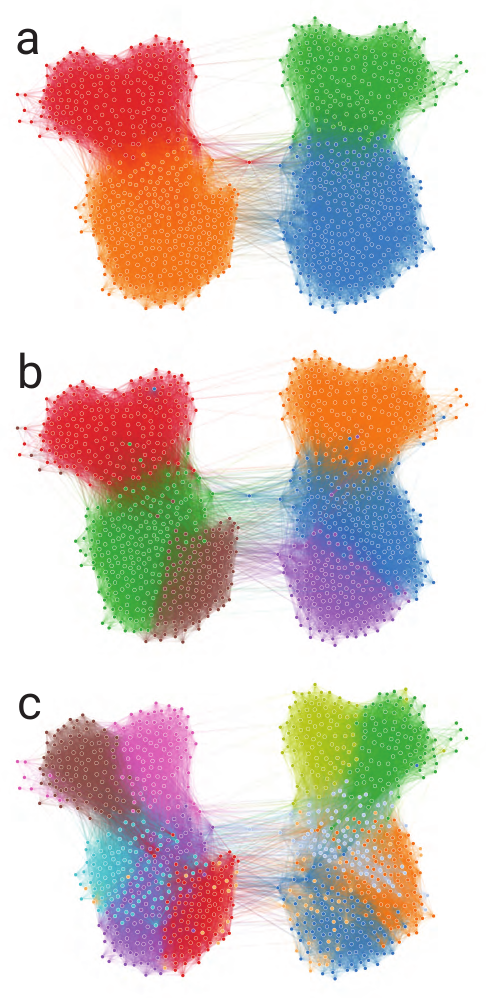}
    \caption{Visualization of the communities obtained for the male 20k network from the Budapest dataset 
    using (a) UL, (b) CL with the range [43,187], (c) CL with the arbitrary range [50,100]. 
    }
\label{fig:brainviz}
\end{figure}

We move on to real networks to demonstrate the applicability of the proposed CL method. The Budapest brain connectome network is a parameterizable consensus network derived from MRI data on brain connectomes \cite{Budapest2015}. In this network, anatomical brain regions are represented as 1015 nodes while edges are created based on an aggregate of several measurements \cite{Budapest2015}. 
Applying UL on this brain network leads to a trivial partition of the brain regions into four communities shown in Figure \ref{fig:brainviz}(a). The four communities obtained this way often correspond to the front-posterior parts of the two hemispheres which are not particularly informative. 

Multiple well studied brain parcellation atlases divide the brain connectome into clusters of nodes \cite{sevenyeo2011organization,atlasfan2016human,atlasschaefer2018local} based on cognitive and brain functions. The relative size of these clusters can inform the user of CL when analyzing the Budapest brain network. 
To obtain communities aligned with exogenous neuroscience knowledge, the subject matter expert may decide to use the community size range of $[43, 187]$ (see Appendix B for the detailed calculations) which is based on the average of relative cluster sizes from the Brainnetome atlas \cite{atlasfan2016human} and the Schaefer atlas \cite{atlasschaefer2018local}.
Given this range, the CL method returns the partition shown in Figure \ref{fig:brainviz}(b) that has 6 communities. The number of communities is more aligned with the 7 well known functionally coupled brain clusters introduced by \citet{sevenyeo2011organization} that are also considered in the two brain atlases \cite{atlasfan2016human,atlasschaefer2018local}. Using the community size range $[50,100]$ will lead to 10 communities (Figure \ref{fig:brainviz}(c)) which is not particularly aligned with neuroscience knowledge. This example demonstrates how CL combined with a size range that is supported by the subject matter expertise allows users to obtain relevant and informative clusters that are aligned with domain-specific knowledge.

\begin{figure}[htbp!]
    \centering
    \includegraphics[width=0.98\linewidth]{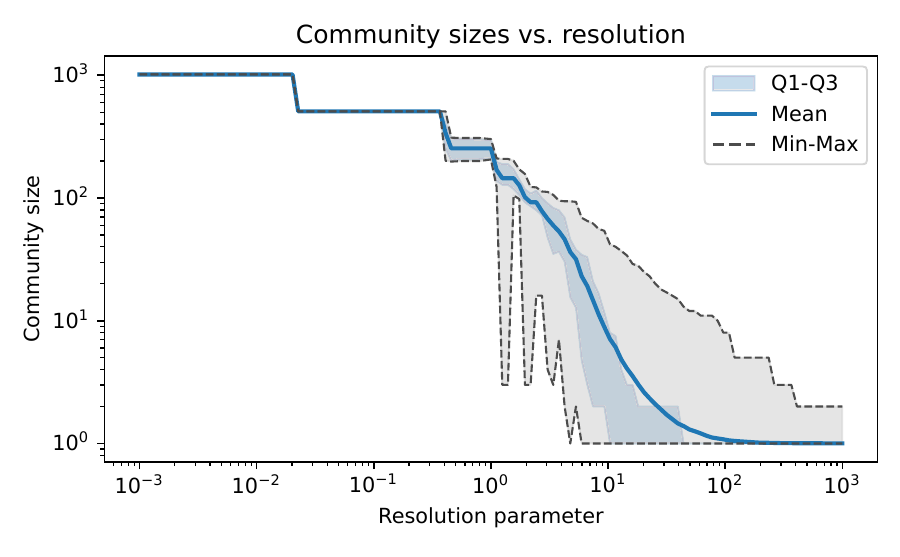}
    \caption{Detected community sizes as a function of the resolution parameter using UL. The mean community size is shown as a blue line, with the shaded region indicating the $25$–$75\%$ quantiles. The gray area with dashed boundaries represents the minimum and maximum community sizes.}
\label{fig:brainstats}
\end{figure}

\subsection{Direct control of community sizes and its variation}

The issue of obtaining communities that are inconsistent with subject matter expertise had been conventionally addressed by changing the resolution parameter. However, one may argue that expressing the subject matter expertise via the community size range is often more straightforward than the ad-hoc practice of trying many different resolution values until arriving at a partition that is aligned with the subject matter expertise knowledge. More importantly, changing the resolution parameter indirectly controls the average community size while it does not control the variation in the community sizes. Figure \ref{fig:brainstats} shows three statistics for community sizes as a function of the resolution parameter when UL is applied on the same Budapest brain network of Figure \ref{fig:brainviz}. Except for resolution values smaller than 0.3 (which lead to a left-right bi-partition or a single-community partition), there can be considerable variation in community sizes. Importantly, the variation in community sizes cannot be controlled through the resolution parameter. For example, Figure \ref{fig:brainstats} shows that, when the chosen resolution value is $\gamma=0.4$, the Q1 and Q3 quartiles for community sizes are roughly 200 and 500 nodes. For larger values of $\gamma$, the variation in community sizes is even larger. The resolution parameter does not control the variation, while using CL offers that option to the user.

To summarize, in the context where the user seeks communities under a restricted community size range, the CL method offers a functionality that is more mathematically grounded (explicit modeling of community sizes) and more computationally efficient (single run) compared to the currently common practice of changing the resolution parameter. More importantly, CL offers direct control of the community size variation that is not achievable through changing the resolution parameter.  

\section{Conclusion}

Community detection is a common task in analyzing complex networks in various domains.
Across these domains there is sometimes prior information available about reasonable community sizes.
In traditional community detection methods, this prior information is not used directly, despite it may help find more informative partitions aligned with subject matter knowledge.

In this article we proposed a modularity-based heuristic that considers such prior information by setting explicit community-size constraints.
We have implemented these constraints in the heuristic Leiden algorithm and compared it to an exact size-constrained modularity optimization baseline.
Both heuristic and exact approaches showed solid performance in finding partitions satisfying the community size constraints. 
While our heuristic results on small instances are on par with our exact global optimization results, the constrained Leiden heuristic is scalable (unlike the exact method) and therefore it can also be applied to large networks.
We demonstrated through six experiments how our methods is advantageous with respect to the common practice of changing resolution parameter values, when a community size range is desired.


We stress that the approach presented in this paper can be applied to the constrained optimization of other quality functions, such as that of the Constant Potts Model (CPM)~\cite{traag11}. Having the possibility to inform community detection by explicitly setting some community size constraints help users incorporate subject matter expertise in a principled way and control both community sizes and the variation among community sizes.

As a future direction, one may consider that modularity suffers from well documented drawbacks in terms of interpretation and lacks a clear statistical foundation. Other approaches, such as stochastic block models (SBMs) may offer methods that are more appropriate for inferential statistics. Future research can explore how prior information on community sizes can be exploited in SBMs, and could be included as explicit priors on distributions of community sizes instead of non-informative or uniform priors (see also~\cite{gosgens_detecting_2024}).

\subsection*{Acknowledgements} 
We acknowledge Majid Saberi and Milad Mokhtaridoost for helpful discussions. 

\subsection*{Code availability}

Code for the Leiden algorithm with community size constraints is available from \url{https://github.com/vtraag/leidenalg} for the Python interface, with the underlying implementation in C++ available from \\ \url{https://github.com/vtraag/libleidenalg}.

Code for the analyses and Constrained Integer Programming (CIP) model is available from \\ \url{https://github.com/filipinascimento/constrainedmodularity}.

\section*{Appendix A: Additional results on the planted partition benchmark networks}

\begin{figure*}
    \centering
    \includegraphics[width=0.98\linewidth]{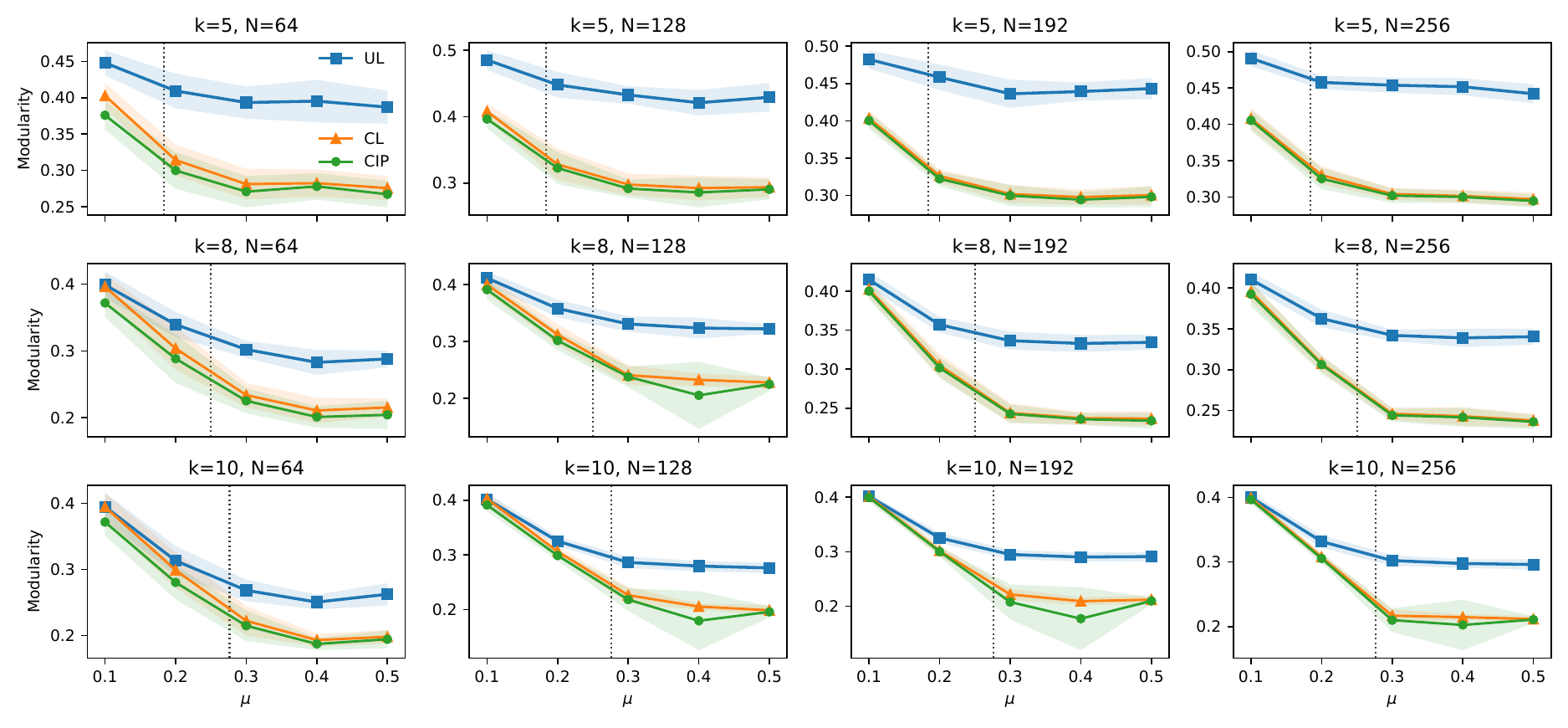}
    \caption{Maximum modularity of partitions obtained by the three methods UL, CL, and CIP for the planted partition networks examined in Section~\ref{sec:modnet}. The vertical dotted line in each panel marks the detectability limit for the corresponding configuration.}
\label{fig:modularity_constrained}
\end{figure*}

Figure \ref{fig:modularity_constrained} shows the maximum modularity values of the partitions obtained by the three methods UL, CL, and CIP for the planted partition networks.

\begin{figure*}
    \centering
    \includegraphics[width=0.98\linewidth]{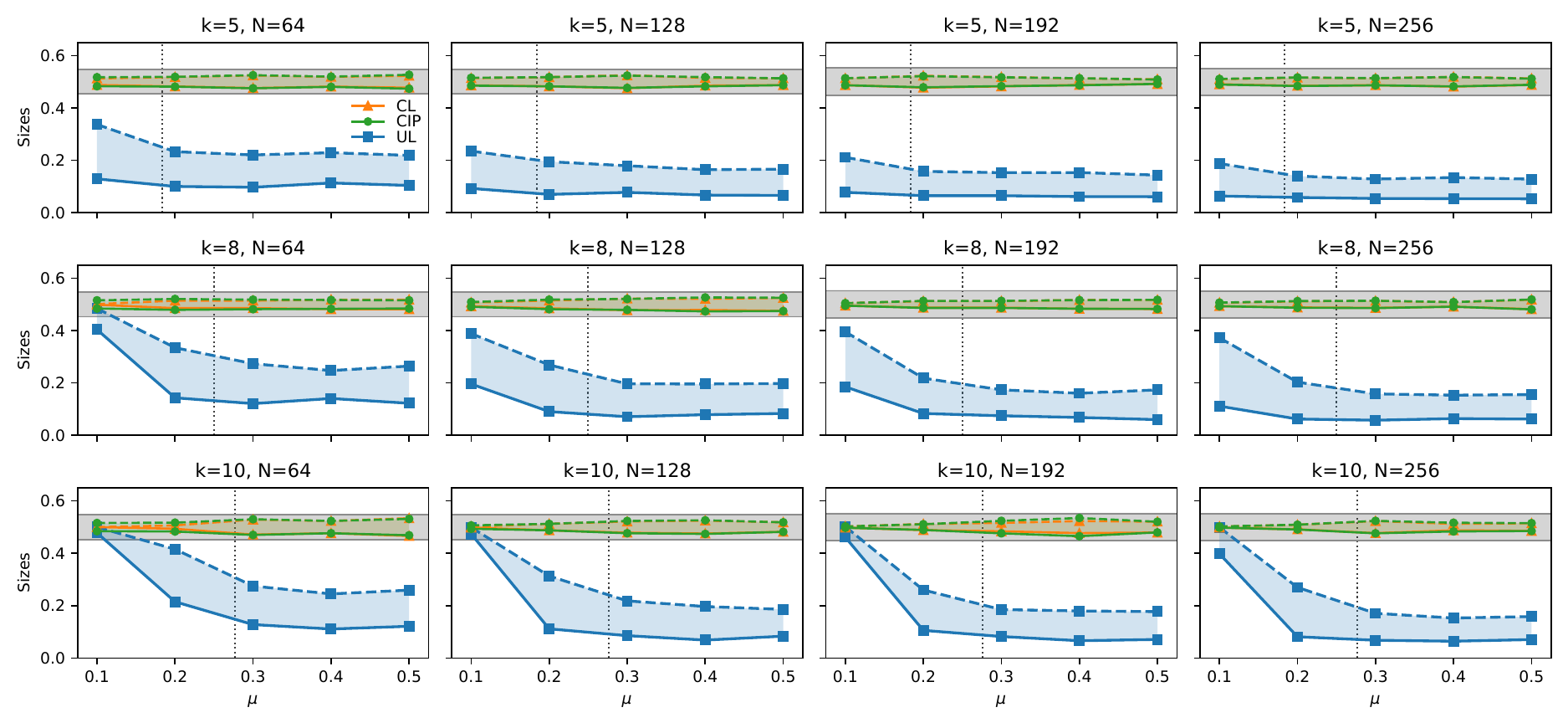}
      \caption{Community sizes from the three methods UL, CL, and CIP for the planted partition networks examined in Section~\ref{sec:modnet}. The grey band shows the expected community size range. The vertical dotted line in each panel marks the detectability limit for the corresponding configuration.}
\label{fig:sizes_constrained}
\end{figure*}

Figure \ref{fig:sizes_constrained} shows the community sizes of the partitions obtained by the three methods UL, CL, and CIP for the planted partition networks.

\section*{Appendix B: Community size range for the brain network aligned with the expert matter expertise}

Multiple well studied brain parcellation atlases divide the brain connectome into clusters of nodes \cite{sevenyeo2011organization,atlasfan2016human,atlasschaefer2018local} based on cognitive and brain functions. The relative size of these clusters can inform the user of CL when analyzing the Budapest brain network. According to the Brainnetome atlas \cite{atlasfan2016human}, the relative minimum and maximum community sizes are 6/246 and 34/246 respectively. These relative community sizes are 12/200 and 46/200, respectively, according to the Schaefer atlas \cite{atlasschaefer2018local}. Using these two sources of exogenous information, the user can provide the minimum community size of $(6/246+12/200)\times1015/2=43$ and the maximum community size of $(34/246+46/200)\times1015/2=187$ nodes. Accordingly, we provided the range $[43, 187]$ to the CL in Section~\ref{sec:brain}.


\begin{thebibliography}{27}%
\makeatletter
\providecommand \@ifxundefined [1]{%
 \@ifx{#1\undefined}
}%
\providecommand \@ifnum [1]{%
 \ifnum #1\expandafter \@firstoftwo
 \else \expandafter \@secondoftwo
 \fi
}%
\providecommand \@ifx [1]{%
 \ifx #1\expandafter \@firstoftwo
 \else \expandafter \@secondoftwo
 \fi
}%
\providecommand \natexlab [1]{#1}%
\providecommand \enquote  [1]{``#1''}%
\providecommand \bibnamefont  [1]{#1}%
\providecommand \bibfnamefont [1]{#1}%
\providecommand \citenamefont [1]{#1}%
\providecommand \href@noop [0]{\@secondoftwo}%
\providecommand \href [0]{\begingroup \@sanitize@url \@href}%
\providecommand \@href[1]{\@@startlink{#1}\@@href}%
\providecommand \@@href[1]{\endgroup#1\@@endlink}%
\providecommand \@sanitize@url [0]{\catcode `\\12\catcode `\$12\catcode
  `\&12\catcode `\#12\catcode `\^12\catcode `\_12\catcode `\%12\relax}%
\providecommand \@@startlink[1]{}%
\providecommand \@@endlink[0]{}%
\providecommand \url  [0]{\begingroup\@sanitize@url \@url }%
\providecommand \@url [1]{\endgroup\@href {#1}{\urlprefix }}%
\providecommand \urlprefix  [0]{URL }%
\providecommand \Eprint [0]{\href }%
\providecommand \doibase [0]{http://dx.doi.org/}%
\providecommand \selectlanguage [0]{\@gobble}%
\providecommand \bibinfo  [0]{\@secondoftwo}%
\providecommand \bibfield  [0]{\@secondoftwo}%
\providecommand \translation [1]{[#1]}%
\providecommand \BibitemOpen [0]{}%
\providecommand \bibitemStop [0]{}%
\providecommand \bibitemNoStop [0]{.\EOS\space}%
\providecommand \EOS [0]{\spacefactor3000\relax}%
\providecommand \BibitemShut  [1]{\csname bibitem#1\endcsname}%
\let\auto@bib@innerbib\@empty
\bibitem [{\citenamefont {Fortunato}(2010)}]{fortunato10}%
  \BibitemOpen
  \bibfield  {author} {\bibinfo {author} {\bibfnamefont {S.}~\bibnamefont
  {Fortunato}},\ }\href@noop {} {\bibfield  {journal} {\bibinfo  {journal}
  {Phys. Rep.}\ }\textbf {\bibinfo {volume} {486}},\ \bibinfo {pages} {75}
  (\bibinfo {year} {2010})}\BibitemShut {NoStop}%
\bibitem [{\citenamefont {Korczak}\ \emph {et~al.}(2019)\citenamefont
  {Korczak}, \citenamefont {Pondel},\ and\ \citenamefont
  {Sroka}}]{korczak2019approach}%
  \BibitemOpen
  \bibfield  {author} {\bibinfo {author} {\bibfnamefont {J.}~\bibnamefont
  {Korczak}}, \bibinfo {author} {\bibfnamefont {M.}~\bibnamefont {Pondel}}, \
  and\ \bibinfo {author} {\bibfnamefont {W.}~\bibnamefont {Sroka}},\ }in\
  \href@noop {} {\emph {\bibinfo {booktitle} {2019 Federated Conference on
  Computer Science and Information Systems (FedCSIS)}}}\ (\bibinfo
  {organization} {IEEE},\ \bibinfo {year} {2019})\ pp.\ \bibinfo {pages}
  {675--683}\BibitemShut {NoStop}%
\bibitem [{\citenamefont {Zhang}\ \emph {et~al.}(2021)\citenamefont {Zhang},
  \citenamefont {Priestley}, \citenamefont {DeMaio}, \citenamefont {Ni},\ and\
  \citenamefont {Tian}}]{zhang2021measuring}%
  \BibitemOpen
  \bibfield  {author} {\bibinfo {author} {\bibfnamefont {L.}~\bibnamefont
  {Zhang}}, \bibinfo {author} {\bibfnamefont {J.}~\bibnamefont {Priestley}},
  \bibinfo {author} {\bibfnamefont {J.}~\bibnamefont {DeMaio}}, \bibinfo
  {author} {\bibfnamefont {S.}~\bibnamefont {Ni}}, \ and\ \bibinfo {author}
  {\bibfnamefont {X.}~\bibnamefont {Tian}},\ }\href@noop {} {\bibfield
  {journal} {\bibinfo  {journal} {Big data}\ }\textbf {\bibinfo {volume} {9}},\
  \bibinfo {pages} {132} (\bibinfo {year} {2021})}\BibitemShut {NoStop}%
\bibitem [{\citenamefont {Szalkai}\ \emph {et~al.}(2015)\citenamefont
  {Szalkai}, \citenamefont {Kerepesi}, \citenamefont {Varga},\ and\
  \citenamefont {Grolmusz}}]{Budapest2015}%
  \BibitemOpen
  \bibfield  {author} {\bibinfo {author} {\bibfnamefont {B.}~\bibnamefont
  {Szalkai}}, \bibinfo {author} {\bibfnamefont {C.}~\bibnamefont {Kerepesi}},
  \bibinfo {author} {\bibfnamefont {B.}~\bibnamefont {Varga}}, \ and\ \bibinfo
  {author} {\bibfnamefont {V.}~\bibnamefont {Grolmusz}},\ }\href {\doibase
  https://doi.org/10.1016/j.neulet.2015.03.071} {\bibfield  {journal} {\bibinfo
   {journal} {Neuroscience Letters}\ }\textbf {\bibinfo {volume} {595}},\
  \bibinfo {pages} {60} (\bibinfo {year} {2015})}\BibitemShut {NoStop}%
\bibitem [{\citenamefont {Mokhtaridoost}\ \emph {et~al.}(2024)\citenamefont
  {Mokhtaridoost}, \citenamefont {Chalmers}, \citenamefont {Soleimanpoor},
  \citenamefont {McMurray}, \citenamefont {Lato}, \citenamefont {Nguyen},
  \citenamefont {Musienko}, \citenamefont {Nash}, \citenamefont {Espeso-Gil},
  \citenamefont {Ahmed}, \citenamefont {Delfosse}, \citenamefont {Browning},
  \citenamefont {Barutcu}, \citenamefont {Wilson}, \citenamefont {Liehr},
  \citenamefont {Shlien}, \citenamefont {Aref}, \citenamefont {Joyce},
  \citenamefont {Weise},\ and\ \citenamefont {Maass}}]{mokhtaridoost2024}%
  \BibitemOpen
  \bibfield  {author} {\bibinfo {author} {\bibfnamefont {M.}~\bibnamefont
  {Mokhtaridoost}}, \bibinfo {author} {\bibfnamefont {J.~J.}\ \bibnamefont
  {Chalmers}}, \bibinfo {author} {\bibfnamefont {M.}~\bibnamefont
  {Soleimanpoor}}, \bibinfo {author} {\bibfnamefont {B.~J.}\ \bibnamefont
  {McMurray}}, \bibinfo {author} {\bibfnamefont {D.~F.}\ \bibnamefont {Lato}},
  \bibinfo {author} {\bibfnamefont {S.~C.}\ \bibnamefont {Nguyen}}, \bibinfo
  {author} {\bibfnamefont {V.}~\bibnamefont {Musienko}}, \bibinfo {author}
  {\bibfnamefont {J.~O.}\ \bibnamefont {Nash}}, \bibinfo {author}
  {\bibfnamefont {S.}~\bibnamefont {Espeso-Gil}}, \bibinfo {author}
  {\bibfnamefont {S.}~\bibnamefont {Ahmed}}, \bibinfo {author} {\bibfnamefont
  {K.}~\bibnamefont {Delfosse}}, \bibinfo {author} {\bibfnamefont {J.~W.~L.}\
  \bibnamefont {Browning}}, \bibinfo {author} {\bibfnamefont {A.~R.}\
  \bibnamefont {Barutcu}}, \bibinfo {author} {\bibfnamefont {M.~D.}\
  \bibnamefont {Wilson}}, \bibinfo {author} {\bibfnamefont {T.}~\bibnamefont
  {Liehr}}, \bibinfo {author} {\bibfnamefont {A.}~\bibnamefont {Shlien}},
  \bibinfo {author} {\bibfnamefont {S.}~\bibnamefont {Aref}}, \bibinfo {author}
  {\bibfnamefont {E.~F.}\ \bibnamefont {Joyce}}, \bibinfo {author}
  {\bibfnamefont {A.}~\bibnamefont {Weise}}, \ and\ \bibinfo {author}
  {\bibfnamefont {P.~G.}\ \bibnamefont {Maass}},\ }\href {\doibase
  10.1038/s41467-024-53983-y} {\bibfield  {journal} {\bibinfo  {journal}
  {Nature Communications}\ }\textbf {\bibinfo {volume} {15}},\ \bibinfo {pages}
  {9813} (\bibinfo {year} {2024})}\BibitemShut {NoStop}%
\bibitem [{\citenamefont {Quinodoz}\ \emph {et~al.}(2018)\citenamefont
  {Quinodoz}, \citenamefont {Ollikainen}, \citenamefont {Tabak}, \citenamefont
  {Palla}, \citenamefont {Schmidt}, \citenamefont {Detmar}, \citenamefont
  {Lai}, \citenamefont {Shishkin}, \citenamefont {Bhat}, \citenamefont {Takei}
  \emph {et~al.}}]{chromosomal}%
  \BibitemOpen
  \bibfield  {author} {\bibinfo {author} {\bibfnamefont {S.~A.}\ \bibnamefont
  {Quinodoz}}, \bibinfo {author} {\bibfnamefont {N.}~\bibnamefont
  {Ollikainen}}, \bibinfo {author} {\bibfnamefont {B.}~\bibnamefont {Tabak}},
  \bibinfo {author} {\bibfnamefont {A.}~\bibnamefont {Palla}}, \bibinfo
  {author} {\bibfnamefont {J.~M.}\ \bibnamefont {Schmidt}}, \bibinfo {author}
  {\bibfnamefont {E.}~\bibnamefont {Detmar}}, \bibinfo {author} {\bibfnamefont
  {M.~M.}\ \bibnamefont {Lai}}, \bibinfo {author} {\bibfnamefont {A.~A.}\
  \bibnamefont {Shishkin}}, \bibinfo {author} {\bibfnamefont {P.}~\bibnamefont
  {Bhat}}, \bibinfo {author} {\bibfnamefont {Y.}~\bibnamefont {Takei}},  \emph
  {et~al.},\ }\href@noop {} {\bibfield  {journal} {\bibinfo  {journal} {Cell}\
  }\textbf {\bibinfo {volume} {174}},\ \bibinfo {pages} {744} (\bibinfo {year}
  {2018})}\BibitemShut {NoStop}%
\bibitem [{\citenamefont {Dekker}\ \emph {et~al.}(2023)\citenamefont {Dekker},
  \citenamefont {Alber}, \citenamefont {Aufmkolk}, \citenamefont {Beliveau},
  \citenamefont {Bruneau}, \citenamefont {Belmont}, \citenamefont {Bintu},
  \citenamefont {Boettiger}, \citenamefont {Calandrelli}, \citenamefont
  {Disteche} \emph {et~al.}}]{topologically}%
  \BibitemOpen
  \bibfield  {author} {\bibinfo {author} {\bibfnamefont {J.}~\bibnamefont
  {Dekker}}, \bibinfo {author} {\bibfnamefont {F.}~\bibnamefont {Alber}},
  \bibinfo {author} {\bibfnamefont {S.}~\bibnamefont {Aufmkolk}}, \bibinfo
  {author} {\bibfnamefont {B.~J.}\ \bibnamefont {Beliveau}}, \bibinfo {author}
  {\bibfnamefont {B.~G.}\ \bibnamefont {Bruneau}}, \bibinfo {author}
  {\bibfnamefont {A.~S.}\ \bibnamefont {Belmont}}, \bibinfo {author}
  {\bibfnamefont {L.}~\bibnamefont {Bintu}}, \bibinfo {author} {\bibfnamefont
  {A.}~\bibnamefont {Boettiger}}, \bibinfo {author} {\bibfnamefont
  {R.}~\bibnamefont {Calandrelli}}, \bibinfo {author} {\bibfnamefont {C.~M.}\
  \bibnamefont {Disteche}},  \emph {et~al.},\ }\href@noop {} {\bibfield
  {journal} {\bibinfo  {journal} {Molecular cell}\ }\textbf {\bibinfo {volume}
  {83}},\ \bibinfo {pages} {2624} (\bibinfo {year} {2023})}\BibitemShut
  {NoStop}%
\bibitem [{\citenamefont {Newman}\ and\ \citenamefont
  {Girvan}(2004)}]{newman_finding_2004}%
  \BibitemOpen
  \bibfield  {author} {\bibinfo {author} {\bibfnamefont {M.~E.~J.}\
  \bibnamefont {Newman}}\ and\ \bibinfo {author} {\bibfnamefont
  {M.}~\bibnamefont {Girvan}},\ }\href {\doibase 10.1103/PhysRevE.69.026113}
  {\bibfield  {journal} {\bibinfo  {journal} {Physical Review E}\ }\textbf
  {\bibinfo {volume} {69}},\ \bibinfo {pages} {026113} (\bibinfo {year}
  {2004})}\BibitemShut {NoStop}%
\bibitem [{\citenamefont {Aref}\ and\ \citenamefont
  {Mostajabdaveh}(2024)}]{aref2024analyzing}%
  \BibitemOpen
  \bibfield  {author} {\bibinfo {author} {\bibfnamefont {S.}~\bibnamefont
  {Aref}}\ and\ \bibinfo {author} {\bibfnamefont {M.}~\bibnamefont
  {Mostajabdaveh}},\ }\href@noop {} {\bibfield  {journal} {\bibinfo  {journal}
  {Journal of Computational Science}\ }\textbf {\bibinfo {volume} {78}},\
  \bibinfo {pages} {102283} (\bibinfo {year} {2024})}\BibitemShut {NoStop}%
\bibitem [{\citenamefont {Newman}(2006)}]{newman_modularity_2006}%
  \BibitemOpen
  \bibfield  {author} {\bibinfo {author} {\bibfnamefont {M.~E.~J.}\
  \bibnamefont {Newman}},\ }\href {\doibase 10.1073/pnas.0601602103} {\bibfield
   {journal} {\bibinfo  {journal} {Proceedings of the National Academy of
  Sciences}\ }\textbf {\bibinfo {volume} {103}},\ \bibinfo {pages} {8577}
  (\bibinfo {year} {2006})}\BibitemShut {NoStop}%
\bibitem [{\citenamefont {Fortunato}\ and\ \citenamefont
  {Hric}(2016)}]{fortunato2016}%
  \BibitemOpen
  \bibfield  {author} {\bibinfo {author} {\bibfnamefont {S.}~\bibnamefont
  {Fortunato}}\ and\ \bibinfo {author} {\bibfnamefont {D.}~\bibnamefont
  {Hric}},\ }\href {\doibase 10.1016/j.physrep.2016.09.002} {\bibfield
  {journal} {\bibinfo  {journal} {Physics Reports}\ }\textbf {\bibinfo {volume}
  {659}},\ \bibinfo {pages} {1} (\bibinfo {year} {2016})}\BibitemShut {NoStop}%
\bibitem [{\citenamefont {Brandes}\ \emph {et~al.}(2007)\citenamefont
  {Brandes}, \citenamefont {Delling}, \citenamefont {Gaertler}, \citenamefont
  {Gorke}, \citenamefont {Hoefer}, \citenamefont {Nikoloski},\ and\
  \citenamefont {Wagner}}]{brandes2007modularity}%
  \BibitemOpen
  \bibfield  {author} {\bibinfo {author} {\bibfnamefont {U.}~\bibnamefont
  {Brandes}}, \bibinfo {author} {\bibfnamefont {D.}~\bibnamefont {Delling}},
  \bibinfo {author} {\bibfnamefont {M.}~\bibnamefont {Gaertler}}, \bibinfo
  {author} {\bibfnamefont {R.}~\bibnamefont {Gorke}}, \bibinfo {author}
  {\bibfnamefont {M.}~\bibnamefont {Hoefer}}, \bibinfo {author} {\bibfnamefont
  {Z.}~\bibnamefont {Nikoloski}}, \ and\ \bibinfo {author} {\bibfnamefont
  {D.}~\bibnamefont {Wagner}},\ }\href@noop {} {\bibfield  {journal} {\bibinfo
  {journal} {IEEE Transactions on Knowledge and Data Engineering}\ }\textbf
  {\bibinfo {volume} {20}},\ \bibinfo {pages} {172} (\bibinfo {year}
  {2007})}\BibitemShut {NoStop}%
\bibitem [{\citenamefont {Agarwal}\ and\ \citenamefont
  {Kempe}(2008)}]{agarwal_modularity-maximizing_2008}%
  \BibitemOpen
  \bibfield  {author} {\bibinfo {author} {\bibfnamefont {G.}~\bibnamefont
  {Agarwal}}\ and\ \bibinfo {author} {\bibfnamefont {D.}~\bibnamefont
  {Kempe}},\ }\href {\doibase 10.1140/epjb/e2008-00425-1} {\bibfield  {journal}
  {\bibinfo  {journal} {The European Physical Journal B}\ }\textbf {\bibinfo
  {volume} {66}},\ \bibinfo {pages} {409} (\bibinfo {year} {2008})}\BibitemShut
  {NoStop}%
\bibitem [{\citenamefont {Dinh}\ and\ \citenamefont
  {Thai}(2015)}]{dinh_toward_2015}%
  \BibitemOpen
  \bibfield  {author} {\bibinfo {author} {\bibfnamefont {T.~N.}\ \bibnamefont
  {Dinh}}\ and\ \bibinfo {author} {\bibfnamefont {M.~T.}\ \bibnamefont
  {Thai}},\ }\href@noop {} {\bibfield  {journal} {\bibinfo  {journal} {Internet
  Mathematics}\ }\textbf {\bibinfo {volume} {11}},\ \bibinfo {pages} {181}
  (\bibinfo {year} {2015})}\BibitemShut {NoStop}%
\bibitem [{\citenamefont {Aref}\ \emph {et~al.}(2023)\citenamefont {Aref},
  \citenamefont {Mostajabdaveh},\ and\ \citenamefont
  {Chheda}}]{aref2023suboptimality}%
  \BibitemOpen
  \bibfield  {author} {\bibinfo {author} {\bibfnamefont {S.}~\bibnamefont
  {Aref}}, \bibinfo {author} {\bibfnamefont {M.}~\bibnamefont {Mostajabdaveh}},
  \ and\ \bibinfo {author} {\bibfnamefont {H.}~\bibnamefont {Chheda}},\ }in\
  \href@noop {} {\emph {\bibinfo {booktitle} {Computational Science -- ICCS
  2023}}},\ \bibinfo {editor} {edited by\ \bibinfo {editor} {\bibfnamefont
  {J.}~\bibnamefont {Miky{\v{s}}ka}}, \bibinfo {editor} {\bibfnamefont
  {C.}~\bibnamefont {de~Mulatier}}, \bibinfo {editor} {\bibfnamefont
  {M.}~\bibnamefont {Paszynski}}, \bibinfo {editor} {\bibfnamefont {V.~V.}\
  \bibnamefont {Krzhizhanovskaya}}, \bibinfo {editor} {\bibfnamefont {J.~J.}\
  \bibnamefont {Dongarra}}, \ and\ \bibinfo {editor} {\bibfnamefont {P.~M.}\
  \bibnamefont {Sloot}}}\ (\bibinfo  {publisher} {Springer Nature
  Switzerland},\ \bibinfo {address} {Cham},\ \bibinfo {year} {2023})\ pp.\
  \bibinfo {pages} {612--626}\BibitemShut {NoStop}%
\bibitem [{\citenamefont {Aref}\ \emph {et~al.}(2024)\citenamefont {Aref},
  \citenamefont {Mostajabdaveh},\ and\ \citenamefont {Chheda}}]{aref2022bayan}%
  \BibitemOpen
  \bibfield  {author} {\bibinfo {author} {\bibfnamefont {S.}~\bibnamefont
  {Aref}}, \bibinfo {author} {\bibfnamefont {M.}~\bibnamefont {Mostajabdaveh}},
  \ and\ \bibinfo {author} {\bibfnamefont {H.}~\bibnamefont {Chheda}},\ }\href
  {\doibase 10.1103/PhysRevE.110.044315} {\bibfield  {journal} {\bibinfo
  {journal} {Physical Review E}\ }\textbf {\bibinfo {volume} {110}},\ \bibinfo
  {pages} {044315} (\bibinfo {year} {2024})}\BibitemShut {NoStop}%
\bibitem [{\citenamefont {{Gurobi Optimization Inc.}}(2026)}]{gurobi}%
  \BibitemOpen
  \bibfield  {author} {\bibinfo {author} {\bibnamefont {{Gurobi Optimization
  Inc.}}},\ }\href@noop {} {\enquote {\bibinfo {title} {Gurobi optimizer
  reference manual},}\ } (\bibinfo {year} {2026}),\ \bibinfo {note} {url:
  \url{https://docs.gurobi.com/current/index.html} date accessed 17 Apr
  2026}\BibitemShut {NoStop}%
\bibitem [{\citenamefont {Karmarkar}(1984)}]{karmarkar1984new}%
  \BibitemOpen
  \bibfield  {author} {\bibinfo {author} {\bibfnamefont {N.}~\bibnamefont
  {Karmarkar}},\ }in\ \href@noop {} {\emph {\bibinfo {booktitle} {Proceedings
  of the sixteenth annual ACM Symposium on Theory of Computing}}}\ (\bibinfo
  {year} {1984})\ pp.\ \bibinfo {pages} {302--311}\BibitemShut {NoStop}%
\bibitem [{\citenamefont {Vinh}\ \emph {et~al.}(2010)\citenamefont {Vinh},
  \citenamefont {Epps},\ and\ \citenamefont {Bailey}}]{vinh_AMI}%
  \BibitemOpen
  \bibfield  {author} {\bibinfo {author} {\bibfnamefont {N.~X.}\ \bibnamefont
  {Vinh}}, \bibinfo {author} {\bibfnamefont {J.}~\bibnamefont {Epps}}, \ and\
  \bibinfo {author} {\bibfnamefont {J.}~\bibnamefont {Bailey}},\ }\href@noop {}
  {\bibfield  {journal} {\bibinfo  {journal} {Journal of Machine Learning
  Research}\ }\textbf {\bibinfo {volume} {11}},\ \bibinfo {pages} {2837}
  (\bibinfo {year} {2010})}\BibitemShut {NoStop}%
\bibitem [{\citenamefont {Jerdee}\ \emph {et~al.}(2025)\citenamefont {Jerdee},
  \citenamefont {Kirkley},\ and\ \citenamefont
  {Newman}}]{jerdee2025normalized}%
  \BibitemOpen
  \bibfield  {author} {\bibinfo {author} {\bibfnamefont {M.}~\bibnamefont
  {Jerdee}}, \bibinfo {author} {\bibfnamefont {A.}~\bibnamefont {Kirkley}}, \
  and\ \bibinfo {author} {\bibfnamefont {M.}~\bibnamefont {Newman}},\
  }\href@noop {} {\bibfield  {journal} {\bibinfo  {journal} {Nature
  Communications}\ }\textbf {\bibinfo {volume} {16}},\ \bibinfo {pages} {11268}
  (\bibinfo {year} {2025})}\BibitemShut {NoStop}%
\bibitem [{\citenamefont {Condon}\ and\ \citenamefont {Karp}(2001)}]{condon01}%
  \BibitemOpen
  \bibfield  {author} {\bibinfo {author} {\bibfnamefont {A.}~\bibnamefont
  {Condon}}\ and\ \bibinfo {author} {\bibfnamefont {R.~M.}\ \bibnamefont
  {Karp}},\ }\href@noop {} {\bibfield  {journal} {\bibinfo  {journal} {Random
  Struct. Algor.}\ }\textbf {\bibinfo {volume} {18}},\ \bibinfo {pages} {116}
  (\bibinfo {year} {2001})}\BibitemShut {NoStop}%
\bibitem [{\citenamefont {Ghasemian}\ \emph {et~al.}(2016)\citenamefont
  {Ghasemian}, \citenamefont {Zhang}, \citenamefont {Clauset}, \citenamefont
  {Moore},\ and\ \citenamefont {Peel}}]{ghasemian2016detectability}%
  \BibitemOpen
  \bibfield  {author} {\bibinfo {author} {\bibfnamefont {A.}~\bibnamefont
  {Ghasemian}}, \bibinfo {author} {\bibfnamefont {P.}~\bibnamefont {Zhang}},
  \bibinfo {author} {\bibfnamefont {A.}~\bibnamefont {Clauset}}, \bibinfo
  {author} {\bibfnamefont {C.}~\bibnamefont {Moore}}, \ and\ \bibinfo {author}
  {\bibfnamefont {L.}~\bibnamefont {Peel}},\ }\href {\doibase
  10.1103/PhysRevX.6.031005} {\bibfield  {journal} {\bibinfo  {journal} {Phys.
  Rev. X}\ }\textbf {\bibinfo {volume} {6}},\ \bibinfo {pages} {031005}
  (\bibinfo {year} {2016})}\BibitemShut {NoStop}%
\bibitem [{\citenamefont {Yeo}\ \emph {et~al.}(2011)\citenamefont {Yeo},
  \citenamefont {Krienen}, \citenamefont {Sepulcre}, \citenamefont {Sabuncu},
  \citenamefont {Lashkari}, \citenamefont {Hollinshead}, \citenamefont
  {Roffman}, \citenamefont {Smoller}, \citenamefont {Z{\"o}llei}, \citenamefont
  {Polimeni} \emph {et~al.}}]{sevenyeo2011organization}%
  \BibitemOpen
  \bibfield  {author} {\bibinfo {author} {\bibfnamefont {B.~T.}\ \bibnamefont
  {Yeo}}, \bibinfo {author} {\bibfnamefont {F.~M.}\ \bibnamefont {Krienen}},
  \bibinfo {author} {\bibfnamefont {J.}~\bibnamefont {Sepulcre}}, \bibinfo
  {author} {\bibfnamefont {M.~R.}\ \bibnamefont {Sabuncu}}, \bibinfo {author}
  {\bibfnamefont {D.}~\bibnamefont {Lashkari}}, \bibinfo {author}
  {\bibfnamefont {M.}~\bibnamefont {Hollinshead}}, \bibinfo {author}
  {\bibfnamefont {J.~L.}\ \bibnamefont {Roffman}}, \bibinfo {author}
  {\bibfnamefont {J.~W.}\ \bibnamefont {Smoller}}, \bibinfo {author}
  {\bibfnamefont {L.}~\bibnamefont {Z{\"o}llei}}, \bibinfo {author}
  {\bibfnamefont {J.~R.}\ \bibnamefont {Polimeni}},  \emph {et~al.},\
  }\href@noop {} {\bibfield  {journal} {\bibinfo  {journal} {Journal of
  neurophysiology}\ } (\bibinfo {year} {2011})}\BibitemShut {NoStop}%
\bibitem [{\citenamefont {Fan}\ \emph {et~al.}(2016)\citenamefont {Fan},
  \citenamefont {Li}, \citenamefont {Zhuo}, \citenamefont {Zhang},
  \citenamefont {Wang}, \citenamefont {Chen}, \citenamefont {Yang},
  \citenamefont {Chu}, \citenamefont {Xie}, \citenamefont {Laird} \emph
  {et~al.}}]{atlasfan2016human}%
  \BibitemOpen
  \bibfield  {author} {\bibinfo {author} {\bibfnamefont {L.}~\bibnamefont
  {Fan}}, \bibinfo {author} {\bibfnamefont {H.}~\bibnamefont {Li}}, \bibinfo
  {author} {\bibfnamefont {J.}~\bibnamefont {Zhuo}}, \bibinfo {author}
  {\bibfnamefont {Y.}~\bibnamefont {Zhang}}, \bibinfo {author} {\bibfnamefont
  {J.}~\bibnamefont {Wang}}, \bibinfo {author} {\bibfnamefont {L.}~\bibnamefont
  {Chen}}, \bibinfo {author} {\bibfnamefont {Z.}~\bibnamefont {Yang}}, \bibinfo
  {author} {\bibfnamefont {C.}~\bibnamefont {Chu}}, \bibinfo {author}
  {\bibfnamefont {S.}~\bibnamefont {Xie}}, \bibinfo {author} {\bibfnamefont
  {A.~R.}\ \bibnamefont {Laird}},  \emph {et~al.},\ }\href@noop {} {\bibfield
  {journal} {\bibinfo  {journal} {Cerebral cortex}\ }\textbf {\bibinfo {volume}
  {26}},\ \bibinfo {pages} {3508} (\bibinfo {year} {2016})}\BibitemShut
  {NoStop}%
\bibitem [{\citenamefont {Schaefer}\ \emph {et~al.}(2018)\citenamefont
  {Schaefer}, \citenamefont {Kong}, \citenamefont {Gordon}, \citenamefont
  {Laumann}, \citenamefont {Zuo}, \citenamefont {Holmes}, \citenamefont
  {Eickhoff},\ and\ \citenamefont {Yeo}}]{atlasschaefer2018local}%
  \BibitemOpen
  \bibfield  {author} {\bibinfo {author} {\bibfnamefont {A.}~\bibnamefont
  {Schaefer}}, \bibinfo {author} {\bibfnamefont {R.}~\bibnamefont {Kong}},
  \bibinfo {author} {\bibfnamefont {E.~M.}\ \bibnamefont {Gordon}}, \bibinfo
  {author} {\bibfnamefont {T.~O.}\ \bibnamefont {Laumann}}, \bibinfo {author}
  {\bibfnamefont {X.-N.}\ \bibnamefont {Zuo}}, \bibinfo {author} {\bibfnamefont
  {A.~J.}\ \bibnamefont {Holmes}}, \bibinfo {author} {\bibfnamefont {S.~B.}\
  \bibnamefont {Eickhoff}}, \ and\ \bibinfo {author} {\bibfnamefont {B.~T.}\
  \bibnamefont {Yeo}},\ }\href@noop {} {\bibfield  {journal} {\bibinfo
  {journal} {Cerebral cortex}\ }\textbf {\bibinfo {volume} {28}},\ \bibinfo
  {pages} {3095} (\bibinfo {year} {2018})}\BibitemShut {NoStop}%
\bibitem [{\citenamefont {Traag}\ \emph {et~al.}(2011)\citenamefont {Traag},
  \citenamefont {Van~Dooren},\ and\ \citenamefont {Nesterov}}]{traag11}%
  \BibitemOpen
  \bibfield  {author} {\bibinfo {author} {\bibfnamefont {V.~A.}\ \bibnamefont
  {Traag}}, \bibinfo {author} {\bibfnamefont {P.}~\bibnamefont {Van~Dooren}}, \
  and\ \bibinfo {author} {\bibfnamefont {Y.}~\bibnamefont {Nesterov}},\
  }\href@noop {} {\bibfield  {journal} {\bibinfo  {journal} {Phys. Rev. E}\
  }\textbf {\bibinfo {volume} {84}},\ \bibinfo {pages} {016114} (\bibinfo
  {year} {2011})}\BibitemShut {NoStop}%
\bibitem [{\citenamefont {Gösgens}(2024)}]{gosgens_detecting_2024}%
  \BibitemOpen
  \bibfield  {author} {\bibinfo {author} {\bibfnamefont {M.}~\bibnamefont
  {Gösgens}},\ }\emph {\bibinfo {title} {Detecting small and large communities
  in networks}},\ \href {\doibase 10.6100/P57G-PP16} {Ph.D. thesis},\ \bibinfo
  {school} {Eindhoven University of Technology} (\bibinfo {year}
  {2024})\BibitemShut {NoStop}%
\end{thebibliography}
\end{document}